\documentclass[twocolumn, useAMS, usenatbib, usegraphicx]{aastex62}
\usepackage{amssymb}
\usepackage{booktabs}
\newcommand\ApJ{ApJ}

\newcommand\ARAA{ARA\&A}

\newcommand\AnA{A\&A}

\def\alf{Alfv\'en\,}
\def\alfc{Alfv\'enic\,}

\def\sg{\sigma}
\def\bq{\begin{equation}}
\def\eq{\end{equation}}
\let\grad=\nabla
\def\cc{{\mbox{cm}^{-3}}}

\def\v{{{\bf{v}}}}

\def\vB{v_B}
\def\va{v_A}
\def\ocd{\omega_{cd}}

\def\vi{{{\bf{v}}}_i}
\def\vj{{{\bf{v}}}_j}
\def\ve{{{\bf{v}}}_e}
\def\vi{{{\bf{v}}}_i}
\def\vd{{{\bf{v}}}_d}
\def\vdz{{{\bf{v}}}_{d0}}

\def\vn{{{\bf{v}}}_n}
\def\B{{\bf{B}}}

\def\J{{{\bf{J}}}}
\def\E{{{\bf{E}}}}
\def\k{{{\bf{k}}}}

\def\x{{{\bf{x}}}}
\def\cc{{\mbox{cm}^{-3}}}
\def\sigv{<\sigma v>}
\newcommand\cross{{\bf{\times}}}
\def\curl{{\grad \cross}}
\newcommand{\dt}[1]{\frac{d #1}{dt}}
\newcommand{\delt} [1] {\frac{\partial #1}{\partial t}}

\received{August 30, 2018}
\revised{December 18, 2018}
\accepted{January 30, 2019}

\submitjournal{AJ}

\begin{document}

\title{Charged grains and Kelvin-—Helmholtz instability in molecular clouds}

\correspondingauthor{B.P. Pandey}
\email{birendra.pandey@mq.edu.au; sergey.vladimirov@sydney.edu.au}

\author[0000-0001-6568-4309]{B.P. Pandey}
\affil{Department of Physics \& Astronomy \& Research centre for Astronomy, Astrophysics and Astrophotonics,\\
Macquarie University, Sydney 2109, Australia}

\author{S.V. Vladimirov}
\affil{Metamaterials Laboratory, National Research University of Information Technology, Mechanics \& Optics, 
\\St. Petersburg 199034, Russia}
\affil{Joint Institute of High Temperatures, Russian Academy of Sciences, \\
Izhorskaya 13/2, 125412 Moscow, Russia}
\affil{ School of Physics, The University of Sydney, Sydney 2006, Australia}

\begin{abstract}
The presence of dust grains profoundly affects the diffusion of the magnetic field in molecular clouds. When the electrons and ions are well coupled to the magnetic field and charged grains are only indirectly coupled, emergent Hall diffusion may dominate over all the other non—-ideal magnetohydrodynamic (MHD) effects in a partially ionized dusty cloud. The low--frequency, long ($\sim 0.01 -– 1\,$ pc) wavelength dispersive MHD waves will propagate in such a medium with the polarization of the waves determined by the dust charge density or, the dust size distribution. In the presence of shear flows, these waves may become Kelvin—-Helmholtz (KH) unstable with the dust charge density or the grain size distribution operating as a switch to the instability. When Hall diffusion time is long (compared to the time over which waves are sheared), the growth rate of the instability in the presence of sub-\alfc flow increases with the charge number $|Z|$ on the grain, while it is quenched in the presence of \alfc or super-\alfc flows. However, when Hall diffusion is fast, the growth rate of the instability depends on the dust charge only indirectly.
\end{abstract}


\section{Introduction}
Molecular clouds are the sites of star formation, and it is believed that the gravitational collapse of massive cloud complexes spanning across light-years is responsible for the star birth. The largest structures of molecular gas, giant molecular clouds (GMCs; $M \sim 10^{5}\,M_{\sun}\,,\,L \sim 50\,\mbox{pc}$) consist of low--density ($n_n \sim 3 \times 10^2\,\mbox{cm}^{-3}$) neutral gas, mostly a mixture of molecular hydrogen and helium with only about $1\%$ of mass in dust grains. As the regions of modest extinction are also the regions where most of the mass of molecular clouds in our Galaxy is contained, the interstellar radiation field is able to maintain sufficient ionization of the partially ionized gas to enable its coupling to the magnetic field \citep{M89}. The Zeeman splitting of OH lines suggests that the presence of $10\--30\,\mu\mbox{G}$ magnetic fields \citep{H87}. Millimeter and far-infrared observations indicate the presence of large--scale ($0.1–-10\,\mbox{pc}$) ordered fields \citep{THH95}. The presence of a milligauss field over several hundred astronomical units in high--density ($n_n \sim 10^8–-10^{10}\,\mbox{cm}^{-3}$) regions is inferred from the Zeeman splitting of OH and $\mbox{H}_2\,\mbox{O}$ maser lines. Because most of the molecular gas is in a photon--dominated region \citep{HT97}, depending on the level of fractional ionization, various non—-ideal magnetic effects may counteract the gravitational condensation of the gas inside GMC cores ($n_n \sim 10^4 -– 10^5\,\mbox{cm}^{-3},\,L$  of a few parsecs).  

A complex network of faint, narrow, relatively diffuse structures called striations is seen in the low column density part of molecular clouds. Although not the site of star formation, these magnetic field--aligned coherent structures have great importance to star formation. Material gets added from the striations to the filament which in turn feeds this material to the ridges where clustered--star formation is ongoing \citep{A13}. Thus material may be fed to denser filaments through striations. The typical velocities of the material flows into or out of these filaments are $\sim 0.5-—1\,\mbox{km}\,\mbox{s}^{-1}$.

The Taurus molecular cloud ($\sim 140\,\mbox{pc}$) presents the most striking evidence of striations as low--level emission located away from the denser filaments \citep{G08}. Striations are also prominent in {\it Herschel} dust continuum maps \citep{PL13}. Making use of observation data on starlight polarization (which is caused by nonspherical dust grains that are preferentially aligned in such a way that their long axis is perpendicular to the ambient magnetic field), a plane—-of—-sky magnetic map of the northwest part of the Taurus molecular cloud, where striations were observed, was created by  \cite{C11}, who found $B\sim 18\,\mu G$ ($B\sim 28\,\mu G$) after applying the Chandrasekhar—-Fermi \citep{CF53} method and $B\sim 35\,\mu G$ ($B\sim 93\,\mu G$) after using the \cite{H09} method in diffuse (filament) regions of the cloud. Note that such a grain alignment mechanism not only rules out the possibility of grain alignment with a longer axis parallel to the magnetic field but contrary to the observations also assumes that all grains are aligned or, marginally aligned to the magnetic field \citep{L07}. The radiative torque mechanism, which overcomes the shortcoming of classical paramagnetic alignment theory, has emerged as the most viable mechanism of grain alignment \citep{A15, HO16}.  
  
Dynamical processes in star forming clouds are strongly controlled by the coupling of a largely neutral medium to the magnetic field. This coupling is facilitated by frequent collisions between the neutrals and plasma particles. The collisional effect in molecular clouds is manifested through various non–-ideal magnetohydrodynamic (MHD) effects. For example, ambipolar diffusion (applicable to the relatively low--density, high ionization fraction regions of the cloud), which redistributes magnetic flux, occurs due to a relative drift of {\it{frozen-in}} ions against the sea of neutrals. When ions are not {\it{frozen-in}} in the field but collide often enough (over the ion gyration period) with the neutrals, causing a transverse (with respect to the ambient magnetic field)  drift between the electrons and ions, Hall diffusion redistributes the magnetic flux. When electrons are not {\it{frozen-in}} in the magnetic field but frequently (over the electron gyration period) collide with the neutrals, Ohm diffusion (applicable to the high--density, low ionization fraction regions) may dissipate magnetic energy in the cloud.  This description of a non--ideal MHD effect is valid only in a dust--free environment. However, molecular clouds are generally dusty. In fact, charged grains are more numerous than plasma particles in dense ($\gtrsim 10^{9}\,\mbox{cm}^{-3}$) cloud cores \citep{NNU91}. Their presence affects not only the ionization structure of the cloud but also its gas--phase abundances \citep{H97, WN99}. Owing to the low ionization fraction ($10^{-4} - 10^{-7}$), grains are either neutral or carry a $\pm 1-\pm 2$ electronic charge \citep{N80}. Further, grains can couple directly or, indirectly to the magnetic field, which will not only modify the ambipolar time--scale \citep{CM93} but may give rise to the Hall effect \citep{WN99}. 

Star formation is often accompanied by energetic outflows, jets, and winds. For example, molecular outflows in regions of massive--star formation have revealed massive and energetic outflows \citep{C97}. Dusty outflows are observed in some of the starburst galaxies \citep{A99}. Most molecular clouds have supersonic turbulence to boot \citep{ZE74, AM75}. Clearly, the presence of flows and inhomogeneity determines ambient physical conditions in the cloud.  The interaction between jets and interstellar gas may lead to the transport of momentum and energy across the shear layer via for example, the Kelvin-—Helmholtz (KH) instability \citep{B95, HS97, D98, R99, Wt04}. The KH instability might as well be responsible for the entrainment of ambient material in massive outflows \citep{Wt07}. The striations in the diffuse non—star--forming regions of molecular clouds may also be explained by invoking the KH instability \citep{H16}. 

The present analysis investigates the KH instability in a partially ionized dusty medium. On physical grounds we anticipate that the weak magnetization of charged dust, which may manifest as the Hall effect \citep{W04, PV06}, modifies not only the onset condition of the KH instability but also the rate at which fluctuations may grow in the fluid. In order to see this, recall that the KH instability occurs at the interface of the shearing fluids and extracts its energy from the velocity difference across the layer to create rotating structures in between. This basic picture is little changed in the ideal MHD. However, in the Hall MHD, the circularly polarized magnetic fluctuations are related to the fluid vortices \citep{BL13}.  Thus depending on the polarization of waves, magnetic fluctuations may help or hinder the rolling up of the interface.  As a result the onset condition of the KH instability will be affected by the Hall diffusion of the magnetic field \citep{P18}.  We shall see that even in sub--\alfc flows, low--frequency waves can destabilize the vortex sheet with the growth rate dependent on the charge number $|Z|\,$ on the dust.

The basic model is discussed in Sec. II. In section III we shall discuss the wave properties of the dusty medium. In section IV the KH instability is discussed. Possible applications of the results are discussed in section V. A brief summary is presented in section VI.       
\section{Basic model}
We shall assume a weakly ionized medium, consisting mainly of neutral particles with a tiny fraction of electrons, ions, and charged and neutral dust grains. The basic set of equations describing the dynamics of such a system is as follows. The continuity equation is 
\bq 
\frac{\partial \rho_j}{\partial t} + \grad\cdot\left(\rho_j\,\vj\right)=0\,. \label{eq:ccnt} 
\eq 
Here, $\rho_j$ is the mass density, and $\vj$ is the velocity of various constituents. The momentum equation for the electrons and ions are 
\bq 
0 = - q_j\,n_j\,\left(\E' + \frac{\vj\cross \B}{c}\right) -
\rho_j\,\nu_{jn}\,\vj\,. 
\label{eq:em1} 
\eq
Here $ q_j\,n_j\,\left(\E' + \vj\cross \B/c\right)$ is the Lorentz force,  where $\E' = \E + \vn\cross \B / c$ is the electric field in the neutral frame with $\E$ and $\B$ as the electric and magnetic field, respectively,  $n_j$ is the number density, $q_e = \mp e$ and $c$ is the speed of light. 
The momentum equations for the negatively charged (subscript $d$) and neutral (subscript $d0$) dust grains are
\bq 
0 = - \,e\,n_d\,\left(\E' + \frac{\vd\cross \B}{c}\right) -
\rho_d\,\nu_{dn}\,\vd - \rho_d\,\nu_{di}^*\,\left(\vd -– \vdz\right)  \,, 
\label{eq:dc} 
\eq
\bq 
0 = - \,\rho_{d0}\,\nu_{d0n}\,\vdz + \rho_{d0}\,\nu_{d0e}^*\,\left(\vd -– \vdz\right)\,. 
\label{eq:dn} 
\eq
We see from Eq.~(\ref{eq:dn}) that when the neutral grain--electron collisions are more frequent than the neutral grain–-neutral particle collisions, i.e. $\nu_{d0n}\ll\nu_{d0e}^*$, the relative drift between the charged and neutral grain, $\left(\vd-–\vdz\right)$ is negligible. In the opposite $\nu_{d0e}^* \ll \nu_{d0n}$ limit, the velocity of the neutral grains is nearly the same as that of the neutral particles, i.e. $\vdz \sim 0$  \citep{KN00}.
 
The neutral momentum equation is 
\bq 
\rho_n\,\dt{\vn} = - \grad{P} + \displaystyle\sum_{e, i, d}
\rho_j\,\nu_{j n}\, \vj\,. 
\label{eq:nm1} 
\eq 
The inertia and pressure gradient terms in Eqs.~(\ref{eq:em1})–-(\ref{eq:dn}) have been dropped because here we are considering a weakly ionized medium. 

The collision frequency is
\bq 
\nu_{j\, n} \equiv \gamma_{j\, n}\,\rho_n = \frac{\sigv_j}{m_j +
m_n}\,\rho_n\,. 
\label{cf0} 
\eq 
Here $\sigv_j$ is the momentum transfer rate coefficient of the $j^{\mbox{th}}$ particle with the neutrals. The ion--neutral, electron--neutral \citep{D11}, dust--neutral \citep{NU86} and neutral—-uncharged dust (subscript $d0$) collision rate coefficients are 
\begin{eqnarray}
<\sigma\,v>_{in}&=&2\cross 10^{-9}\,\left(\frac{m_H}{m_r}\right)^{1/2}\quad
\mbox{cm}^3\,\mbox{s}^{-1}\,,\nonumber \\
<\sigma\,v>_{en}&=&4.5 \cross
10^{-9}\, T_{30}^{\frac{1}{2}}\quad
\mbox{cm}^3\,\mbox{s}^{-1}\,,\nonumber\\
<\sigma\,v>_{dn}&=&2.2 \cross 10^{-5}\, T_{30}^{\frac{1}{2}}\,a_{0.1}^{2}\quad
\mbox{cm}^3\,\mbox{s}^{-1}\,,
\nonumber\\
<\sigma\,v>_{nd0}&=&3.2\cross 10^{-5}\, T_{30}^{\frac{1}{2}}\,a_{0.1}^{2}\quad
\mbox{cm}^3\,\mbox{s}^{-1}\,.
\label{eq:cf1}
\end{eqnarray}
Here $m_H$ is the mass of the hydrogen atom, $m_r=m_i\,m_n/(m_i+m_n)$ is the reduced mass, $T_{30}$ is the gas temperature and $a_{0.1}$ is the grain radius in units of 30\,K and $0.1\,\mu m$ respectively. Note that the collision rate involving dust increases with $a^2$ reflecting the fact that the collision cross section for large grains is geometric, i.e. $\sigma\propto a^2$. However, for small grains, the rate coefficient is similar to that of the ion—-neutral collision rate. 

Adopting $m_i = 30\,m_p$ for the ion mass and $m_n = 2.33\,m_p$ for the mean neutral mass, where $m_p = 1.67 \times 10^{-24}\,\mbox{g}\,,$ and $m_d =4\,\pi\,a^3\,\mbox{g}$ for dust material density $ 3\,\mbox{g}\,\cc$, the collision frequencies become 
\begin{eqnarray} 
\,\nu_{in} &=& 1.4\times 10^{-10}\,n_n\,\mbox{s}^{-1}\,,\nonumber\\
\,\nu_{en} &=& 4.5\times 10^{-9}\, T_{30}^{\frac{1}{2}}\,n_n\,\mbox{s}^{-1}\,,\nonumber\\
\,\nu_{dn} &=& 6.8\times 10^{-15}\,n_n\, T_{30}^{\frac{1}{2}}\,a_{0.1}^{2}\,\mbox{s}^{-1}\,,
\nonumber\\
\,\nu_{d0n} &=& 9.9\times 10^{-15}\,n_n\, T_{30}^{\frac{1}{2}}\,a_{0.1}^{2}\,\mbox{s}^{-1}\,. 
\label{eq:cf2} 
\end{eqnarray} 
For polycyclic aromatic hydrocarbons (PAH) grains ($\sim 3\times 10^{-4}\,\mu m\,,m_d \sim 3\times 10^{-22}\,\mbox{g}$), the dust-neutral collision frequency $\nu_{dn} \simeq 1.4\times 10^{-9}\,n_n\,\mbox{s}^{-1}$ is slightly higher than the ion--neutral collision frequency. However, for micron--sized grains, the charged grain--neutral collision frequency is much lower than the ion--neutral collision frequency. 

The inelastic ion capture rate by the negatively charged grains and the inelastic electron sticking to the neutral grains are given as \citep{S41,S42}
\begin{eqnarray}
\alpha_{id} &=& \pi\,a^2\,\left(\frac{8\,k_B\,T}{\pi\,m_i}\right)^{1/2}\,
\left[ 1 + \left( \frac{e^2}{a\,k_B T} \right)
\right]\,
P_i\, S_i\,,\nonumber\\
\alpha_{ed0} &=& \pi\,a^2\,\left(\frac{8\,k_B\,T}{\pi\,m_e}\right)^{1/2}\,
P_e\,\, S_e\,,
\end{eqnarray}
where
\begin{eqnarray}
P_i &=&   1 + \left[ \frac{2} {\left(a\,k_B\,T/e^2\right) + 2} 
      \right]^{1/2}\,, \nonumber\\ 
P_e &=&  1 + \left( \frac{\pi\,e^2}{2\,a\,k_B T}\right)^{1/2}\,,
\end{eqnarray}
account for the electrostatic polarization of the grains by the electric field of the approaching charged particles \citep{N60,DS87, PV16}. For $a = 0.1\,\mu m$, $T=30\,K$, the contribution of the polarization factor to the ion--negatively charged dust collision rate is $ P_i \sim 1$. For electron-—neutral dust collision, this factor is $ P_e \sim 4$. On the other hand, for micron--sized grains,   electron--neutral dust polarization factor is an order of magnitude ($ P_e \sim 88$) larger than the ion-–dust factor. Assuming that the sticking probabilities $S_e = S_i = 1$, the inelastic charged dust-–ion $\nu_{di}^* = n_i\,\alpha_{id}$ and electron--neutral dust $\nu_{ed0}^* = n_e\,\alpha_{ed0}$ collision frequencies become
\begin{eqnarray}
\nu_{di}^*&=&5.8\times 10^{-5}\left(X_e + \frac{|Z|\,n_d}{n_n}\right)\,n_n\,a_{0.1}^{2}\,T_{30}^{1/2}\, \left(\frac{P_i}{1.96}\right)\,\mbox{s}^{-1}\,,\nonumber\\
\nu_{d0e}^*&=&4.2\times 10^{-3}\,X_e\,n_n\,a_{0.1}^{2}\,T_{30}^{1/2}\,  \left(\frac{P_e}{3.96}\right)\,\mbox{s}^{-1}\,.
\label{eq:icf}
\end{eqnarray} 
We use the plasma quasineutrality condition $n_i = n_e + |Z|\,n_d$ in $\nu_{di}^*$. Here $X_e=n_e/n_n$ is the fractional ionization of the cloud. As can be seen from Eqs.~(\ref{eq:cf2}) and (\ref{eq:icf}), the inelastic collision frequencies ($\nu_{di}^*\,,\nu_{d0e}^*$) can even dominate the ion–-neutral ($\nu_{in}$)  and electron–-neutral ($\nu_{en}$) collision frequencies in the diffuse region ($X_e \sim 10^{-4}$). It is only when $X_e \lesssim 10^{-6}$, does plasma–-neutral collision dominate over the dust–-plasma collision. Clearly, notwithstanding their minuscule number densities, both plasma–-neutral and dust–-plasma collision frequencies could be equally important in diffuse clouds.
 
A comparison between the charged dust-—neutral collision frequency ($\nu_{dn}$), Eq.~(\ref{eq:cf2}), and the inelastic charged dust-–ion ($\nu_{di}^*$) and neutral dust–-electron ($\nu_{d0e}^*$) collision frequencies, Eq.~(\ref{eq:icf}), gives 
\bq
\left(\frac{\nu_{d0e}^*}{\nu_{dn}}\right) \simeq 10^{11}X_e\left(\frac{P_e}{3.96}\right)\,,\nonumber\\
\left(\frac{\nu_{di}^*}{\nu_{dn}}\right) \simeq 10^{10}X_e\left(\frac{P_i}{1.96}\right)\,,
\label{eq:comf}
\eq 
which suggest that for micron-sized grains inelastic dust--plasma collision will dominate dust--neutral collision not only in the diffuse ($X_e \sim 10^{-4}$) clouds but also in dark cores ($X_e \gtrsim 10^{-7}$). This is valid in the presence of very small grains 
as well unless fractional ionization plummets to below $10^{-8}—-10^{-9}$.  

Given that 
\bq
\left(\frac{\nu_{d0n}}{\nu_{d0e}^*}\right)\simeq 2.36\times 10^{-12}\,X_e^{-1}\,\left(\frac{P_e}{3.96}\right)^{-1}\,,
\label{eq:inN}
\eq
i.e. $\nu_{d0n}\ll \nu_{d0e}^*$ for $X_e\lesssim 10^{-12}$, we infer that the velocity of neutral [Eq.~(\ref{eq:dn})], and charged grains is the same in the diffuse and dark clouds. All in all, the inelastic collision and the ensuing charge fluctuation could be important to the cloud dynamics.   
  
The presence of tiny grains may significantly reduce the level of ionization as these grains not only carry a small amount of charge but also provide a large surface area for the recombination of plasma particles. Therefore, magnetic diffusion is significantly modified in the presence of such grains \citep{Z18}. As the inelastic frequencies $\nu_{di}^*$ and $\nu_{d0e}^*$ are a function of $X_e$, magnetization of the grain will depend on the rate at which charge on the grain fluctuates, i.e., on the inelastic momentum exchange. Thus making use of  Eq.~(\ref{eq:dn}) we can write Eq.~(\ref{eq:dc}), as 
\bq 
0 = - \,e\,n_d\,\left(\E' + \frac{\vd\cross \B}{c}\right) 
- \rho_d\,\nu_{g}\,\vd\,, 
\label{eq:dc1} 
\eq
where
\bq
\nu_{g} = \nu_{dn}\,\Bigg[1+ \left(\frac{\nu_{di}^*}{\nu_{dn}}\right)\frac{1}{\left(1 + \nu_{d0e}^{*}/ \nu_{d0n}\right)}\Bigg]\,.
\label{eq:efc}
\eq
In writing Eq.~(\ref{eq:efc}), we use the relation $\rho_d\,\nu_{di}^{*} = \rho_{d0}\,\nu_{d0e}^{*}$, i.e. the time over which a positively charged ion neutralizes a negatively charged grain is the same as the time over which a neutral grain acquires an electron. Thus the grain charges are in equilibrium. The charge fluctuation on the grain results in an increased coupling between the charged dust and neutral particles as from Eq.~(\ref{eq:inN}) we see that $\nu_{d0e}^{*}/ \nu_{d0n}\gg 1$ and thus $\nu_{g} \approx 2\,\nu_{dn}$. Clearly, over $\nu_{g}^{-1}$, grains have an average charge number $|Z|$.  

In order to describe how well charged particles are coupled to the magnetic field, we shall define the following plasma Hall parameter:
\bq
\beta_j = \frac{\omega_{cj}}{\nu_{jn}}\,,
\eq
which is the ratio of the cyclotron ($\omega_{cj} = q_j\,B/m_j\,c$) and collision ($\nu_{jn}$) frequencies. The non-ideal MHD effect can be quantified in terms of various Hall parameters $\beta_j$ \citep{PW08}.  

The magnetic field scales with the neutral density as \citep{D83, WN99}
\bq
B[\mbox{mG}]= 
n_{6}^{\alpha}\,,
\label{eq:fscl}
\eq
where $n_{q}=n_n/10^q\,\cc$ and $\alpha=0.5$ when $n_6\leqslant 1$ and $\alpha=0.25$ for higher densities. Similarly, the fractional ionization scales as
\bq
X_e= \left\{\begin{array}{ll}
    \frac{10^{-7}}{\sqrt{n_{4}}}\,,\quad \mbox{for}\, n_{4}>1\,\\
    10^{-4}\,,\,\,\quad \mbox{for}\,n_{2}=1\,.
\end{array} \right.
\label{eq:fion}
\eq
With the above scaling, the electron and ion Hall parameters become
\begin{eqnarray}
\beta_e&=&4\times 10^{15}\,\left(\frac{B}{n_n}\right)=4\times 
\left\{\begin{array}{ll}
10^{9}\,n_n^{-1/2}\,,\\
10^{10.5}\,n_n^{-3/4}\,,
\end{array}\right.
\nonumber\\
%
%\bq
\beta_i&=&2.2\times 10^{12}\,\left(\frac{B}{n_n}\right)=2.2\times 
\left\{\begin{array}{ll}
10^{6}\,n_n^{-1/2}\,,\\
10^{7.5}\,n_n^{-3/4}\,,
\end{array}\right.
\label{eq:HallB}
\end{eqnarray}
where the upper and lower values of the Hall parameter $\beta$ correspond to $\alpha=1/2\,$ ($n_{6}\leqslant 1$) and $\alpha=1/4$ ($n_{6}>1$) in Eq.~(\ref{eq:fscl}). 

The dust Hall parameter $\beta_d$ requires the knowledge of $|Z|$ and the dust size distribution. For micron-- and submicron--sized grains, the charge number carried by a grain is \citep{D80, PVS11}
\bq
|Z| \approx \frac{4\,a\,k_B\,T}{e^2}\,, 
\label{eq:lZ}
\eq
whereas for very small grains ($a\,k_B\,T/e^2\lesssim 0.2$) the charge number carried by a grain is \citep{DS87}
\bq
|Z| \approx \frac{1}{1+8.6\times\left(10^8\,a\,T\right)^{-1/2}}\,.
\label{eq:sZ}
\eq
Thus making use of Eq.~(\ref{eq:lZ}) the dust Hall parameter $\beta_d$ for the micron-- and submicron--sized grains can be written as
\bq
\beta_d=6.7\times 10^{7}\,a_{0.1}^{-4}\,T_{30}^{1/2}\,\left(\frac{B}{n_n}\right)\,. 
\eq
Thus we get
\bq
\beta_d=\left\{\begin{array}{ll}
67\, a_{0.1}^{-4}\,T_{30}^{1/2}\,n_n^{-1/2}\,,\quad\quad  n_{6}\lesssim 1\,,\\
2\times 10^{3}\,a_{0.1}^{-4}\,T_{30}^{1/2}n_n^{-3/4}\,,\quad\quad  n_{6}> 1\,.
\end{array}\right.
\label{eq:HallBd}
\eq
Clearly, for $0.1\,\mu m$ grains $\beta_d\gtrsim 1$ in  the diffuse as well as dense regions of the cloud. However, $a\gtrsim 0.28\,\mu m$ grains are unmagnetized ($\beta_d\ll 1$) when $n_6<1$ and magnetized in the denser region. With increasing grain size, grains will remain unmagnetized in the diffuse as well as in the dense region of the cloud \citep{W07, Z16}. Thus in molecular clouds ions and electrons are tied to magnetic fields, whereas the largest grains are not. 

Very small (few nanometers in size) grain, because of their smaller size, they carry less charge than a large grains; electron collide less frequently with the negatively charged spheres of decreasing radii and steepening Coulomb potentials \citep{PC11}. Clearly due to their higher Coulomb potential barrier the average charge fluctuates around $|Z| \approx 1$. Therefore, the Hall $\beta_d$ for a tiny grain is similar to the ion Hall $\beta_i$ as the reduced gyrofrequency due to larger mass (compared to that of ions) is compensated by the reduced momentum transfer rate due to larger inertia \citep{B11}.

In order to find the charge density $|Z|\,n_d$ in the cloud, Eqs.~(\ref{eq:lZ})--(\ref{eq:sZ}) need to be integrated over the grain size distribution. In diffuse interstellar clouds, small grains are abundant, and the MRN power--law size distribution \citep{M77, NF13}
\bq
\frac{d\,n_d}{d\,a}= A\,n_n\,a^{-3.5}\,,\quad a_1<a<a_2\,,
\eq 
with $a_2\approx 0.25\,\mu\,m$ and $a_1\approx 3\,\mbox{\AA}$, is consistent with the observed extinction. After integrating over $a$ between $a_1$ and $a_2$, we have \citep{DS87} 
\bq
\frac{|Z|\,n_d}{n_n}=
\left\{\begin{array}{ll}
   2.3\times 10^{-13}\,\,T_{30}\,\left(\frac{a_1}{0.1\,\mu m}\right)^{-3/2}\,,\\
    3.8\times 10^{-7}\,\left(\frac{ a_1}{3\times 10^{-4}\mu m}\right)^{-5/2}\,.
\end{array} \right.
\label{eq:smg}
\eq
Assuming a dust material density of $ 3\,\mbox{g}\,\mbox{cm}^{-3}$ and interstellar dust abundance $\rho_d/\rho_{n}=0.01$ gives $n_d/n_n=0.01\,(m_n/m_d)$. Thus the average charge for $0.1-10\,\mu m$ grains is $|Z|=0.1-20$ while  for PAH-—like grains $|Z|=.003$. 

Often we need to know the Havnes parameter $|Z|\,n_d/n_e$ and thus it is desirable to express Eq.~(\ref{eq:smg}) in terms of the Havnes parameter: 
\bq
\left(\frac{|Z|\,n_d}{n_e}\right) =
\left\{\begin{array}{ll}
   2.3\times 10^{-13}\,X_e^{-1}\,T_{30}\,\left(\frac{a_1}{0.1\,\mu m}\right)^{-3/2}\,,\\
    3.8\times 10^{-7}\,\left(\frac{ a_1}{3\times 10^{-4}\mu m}\right)^{-5/2}\,.
\end{array} \right.
\label{eq:smg1}
\eq

We shall define the mass density and bulk velocity of the fluid as $ \rho \approx \rho_n$ and $ \v \approx \v_n$.  The continuity equation [summing up equation (\ref{eq:ccnt})] for the bulk fluid becomes 
\bq 
\delt\rho + \grad\cdot\left(\rho\,\v\right) = 0\,. 
\label{eq:cont1} 
\eq 
The momentum equation can be derived by adding equations (\ref{eq:em1})-(\ref{eq:nm1}):
 \bq
\rho\,\frac{d\v}{dt} =  - \nabla\,P + \frac{\J\cross\B}{c}\,.
\label{eq:meq} 
\eq
In the present work we shall assume $\beta_d \ll 1$. The transverse (to the magnetic field) component of the plasma velocities in Eq.~(\ref{eq:em1}) can be written as 
\bq
{\vj}_{\perp}=\frac{\mp \beta_j\,\frac{c\,\E'}{B} + \beta_j^2\,\frac{c\,\E'\cross \B}{B^2}}{1+\beta_j^2}\,,
\label{eq:vper}
\eq
where the minus (plus) sign corresponds to the electrons (ions). 
As $\beta_e \gg \beta_i \gg 1$, this implies that
\bq
{\ve}_{\perp}\simeq {\vi}_{\perp} \approx \frac{c\,\E'\cross \B}{B^2}\,.
\eq 
Thus we shall assume $\ve\simeq\vi$. The resulting current density $\J = e\,\left(n_i - n_e \right)\,\v_e + |Z|\,e\,n_d\,\vd$ becomes
\bq 
\ve = \frac{- \J}{|Z|\,e\,n_d} + \vd\,. 
\label{eq:vdf} 
\eq
Note that ${\vd}_{\perp} \approx -\beta_d\,c\,\E'/B$, and thus with respect to the plasma particles dust is almost immobile because $\beta_d\ll1$.  

Taking the curl of the electron momentum equation (\ref{eq:em1}) and making use of Maxwell's equations, the induction equation can be written as
\bq 
\delt \B = \curl\,\left[\left(\v+\vB\right)\cross\B\right]\,. 
\label{eq:ind} 
\eq
In the above equation
\bq
\vB = -– \eta_H\,\frac{\left(\grad\cross\B\right)_{\perp}}{B}\,, 
\label{eq:md0}
\eq
is the drift velocity of the magnetic field. The Hall diffusion coefficient $\eta_H$ in terms of \alf velocity $\va$ and the dust—-cyclotron frequency $\ocd$,  
\bq
\va^2=\frac{B^2}{4\,\pi\,\rho}\,,\quad \ocd=\frac{|Z|\,e\,B}{m_d\,c}\,,
\eq
can be written as 
\bq
\eta_H= \left(\frac{\rho}{\rho_d}\right)\frac{\va^2}{\ocd}\,.
\eq
The charged grains are responsible for the Hall term in the induction Eq.~(\ref{eq:ind}). The Hall electric field is due to the transverse drift of the plasma particles against almost stationary dust \citep{PV06}. However, the transverse drift of plasma particles is mitigated with increasing $|Z|$. The Hall effect introduces a scale,  
\bq
L_D=\left(\frac{\rho}{\rho_d}\right)\left(\frac{\va}{\ocd}\right)\,,
\label{eq:dhs}
\eq  
in an otherwise scale--free dusty fluid. The Hall scale can also be written as
\bq
L_D=3.5\times 10^{7}\,\left(\frac{|Z|\,n_d}{n_n}\right)^{-1}\,n_n^{-1/2}
\,.
\label{eq:dhs1}
\eq
Here we see that $L_D$ is independent of the magnetic field and depends only on how well the electrons are attached to the dust grain.

The dependence of $L_D$ on fraction ionization can be spelled out explicitly by writing it as $L_D=3.5\times 10^{7}\,\left(|Z|\,n_d/n_e\right)^{-1}\,X_e^{-1}\,n_n^{-1/2}$ which after using Eq.~(\ref{eq:smg1}) becomes 
\bq
L_D \big[\mbox{cm}\big]=\left\{\begin{array}{ll}
1.5\times 10^{20}\,T_{30}\,n_n^{-1/2}\,\left(\frac{a_1}{0.1\,\mu m}\right)^{3/2}\,,\nonumber\\
    10^{14}\, n_n^{-1/2}\,\left(\frac{ a_1}{3\times 10^{-4}\mu m}\right)^{5/2}\,.
\end{array} \right.
\label{eq:dhs2}
\eq   
At densities above $n_n\gtrsim 10^{10}\,\mbox{cm}^{-3}$ where most of the negative charge resides on the grain \citep{NNU91, WN99} Hall diffusion could become important in the presence of submicron--sized grains as $L_D$ is $\sim 0.01-—0.001 \mbox{pc}$. However, in the absence of very small grains ambipolar diffusion is the dominant magnetic diffusion in the partially ionized cloud \citep{Z16}. Although Hall is important in the presence of very small grains \citep{Z18}, as we see from Eq.~(\ref{eq:dhs2}), the Hall effect in this case is important only over the sub—-astronomical unit scale; the smaller the grain, the smaller the Hall scale. This is not surprising as $L_D$ is the scale over which dust remains unmagnetized. 

The ratio between convective and magnetic drift velocities in the induction Eq.~(\ref{eq:ind}) gives
\bq
R_{D}=\left(\frac{v\,L}{\eta_H}\right)\sim \left(\frac{L}{L_D}\right)\,, 
\label{RD}
\eq  
which is very similar to the magnetic Reynolds number. Here $L$ is the characteristic length of the system. Note that $R_D \propto |Z|\,n_d/n_e$ and thus the advection dominates Hall when $ Z|\,n_d \gg n_e$. 

Together with the Ampere{}\'s law,
\bq
\J = \frac{c}{4\,\pi}\curl\B\,,
\eq
and an isothermal equation of state $P = C_s^2\,\rho$, Eqs.~(\ref{eq:cont1}), (\ref{eq:meq}) and (\ref{eq:ind}) form the required set of equations. 

\section{Waves in the Medium}
The waves in molecular clouds both with \citep{PHHM87, WN99} and without dust \citep{M11} have been investigated in the past. 
The presence of Hall makes \alf waves dispersive \citep{PW08}. As the next section on the KH instability will require some appreciation of the wave properties in a dusty medium, here we briefly describe low--frequency waves with particular focus on the role of dust charge $|Z|$ in polarizing the waves. After linearizing equations (\ref{eq:cont1}), (\ref{eq:meq}), and (\ref{eq:ind}) against a homogeneous background, we have  
\bq
\left(\partial_t + \v\cdot\nabla\right)\,\delta\rho + \nabla\cdot \left(\rho\,\delta \v \right) = 0\,, 
\label{eq:Lc}
\eq
\bq
\rho\,\left(\partial_t + \v\cdot\nabla\right)\,\delta\v = -\nabla\delta P + \frac{\delta J\cross \B}{c}\,,
\label{eq:Lm}
\eq
\bq
\left(\partial_t + \v\cdot\nabla\right)\,\delta \B = \nabla\cross\big[\left(\delta\v + \delta\v_B\right)\cross\B\big]\,.
\label{eq:Lnd}
\eq
Assuming $\delta f \sim \exp \left(\omega\,t+ i\,\k\cdot\x\right)$ where $\omega$ is the angular frequency and $\k$ is the wave vector, in the absence of flow ($\v = 0$) we get the following dispersion relation: 
\begin{eqnarray}
\left(\omega^2+\omega_A^2\,\cos^2\theta\right)^2+\left(\frac{\omega^2+\omega_A^2\,\cos^2\theta}{\omega^2+k^2\,c_s^2}\right)\times  
\nonumber \\\omega^2\,\omega_A^2\,\sin^2\theta +\left(k^2\,\eta_H\right)^2\,\omega^2\,\cos^2\theta=0\,.
\label{eq:d1f}
\end{eqnarray}
Here $\theta={\bf{k}\cdot\B}/\left(k\,B\right)$ and $\omega_A=k\,\va$. For waves propagating transversely to the ambient field, i.e., $\theta = \pi/2$, the dispersion relation (\ref{eq:d1f}) gives the usual magnetosonic branch:
 \bq
\omega^2 = -k^2\,\left(c_s^2 + \va^2\right)\,.
\label{eq:ms1}
\eq 
For waves propagating along the background magnetic field ($\theta = 0$) the dispersion relation, Eq.~(\ref{eq:d1f}), gives the following roots:
\bq
\omega = \left(\frac{i\,\omega_W}{2}\right)\left(
1 \pm \sqrt{1+4\,\frac{\omega_A^2}{\omega_W^2}}\right)\,.
\label{eq:whist}
\eq
where
\bq
\omega_W = k^2\,\eta_H\equiv \left(k\,L_D\right)\,\omega_A\,, 
\eq
is the whistler frequency.  The positive/negative sign inside the bracket in (\ref{eq:whist}) corresponds to the left/right circularly polarized waves.

\begin{figure}
\includegraphics[scale=0.45]{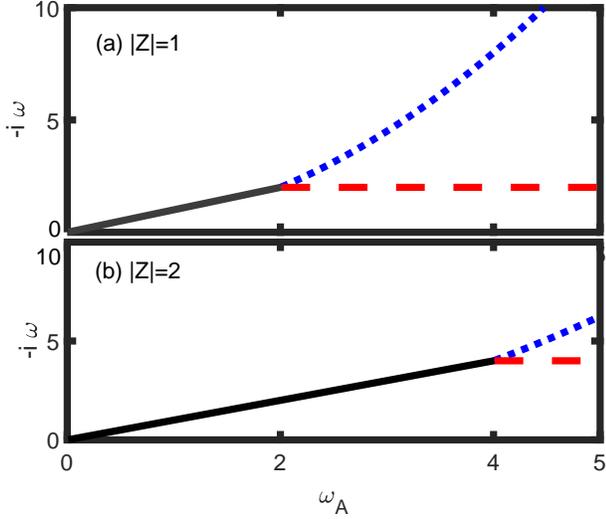}
     \caption{A sketch of the wave propagation in a partially ionized and magnetized dusty plasma is shown in the above figures. The plots of $-i\,\omega$ vs. $\omega_A$ show representative curves corresponding to the \alf (solid), whistler (dotted), and dust cyclotron (dashed) waves for dust carrying a charge number $|Z|=1$ [panel (a)]  and $|Z|=2$ [panel (b)].}
 \label{fig:Fx1}  
\end{figure} 

We see from Eq.~(\ref{eq:whist}) that the high--frequency whistler ($\omega_A \ll \omega$), 
\bq
\left(\frac{\omega}{\omega_A}\right) = i\,\left(k\,L_D\right) \propto \left(\frac{|Z|\,n_d}{n_n}\right)^{-1}\,,
\label{eq:wlr}
\eq
and low--frequency dust cyclotron ($\omega \ll \omega_A$), 
\bq
\left(\frac{\omega}{\omega_A}\right) = \frac{i}{\left(k\,L_D\right)}\propto \left(\frac{|Z|\,n_d}{n_n}\right) \,,
\label{eq:icw}
\eq
are the normal modes in a dusty fluid. Note that the dust cyclotron mode is independent of the wavelength and has been written in the above form to highlight that its frequency is much smaller than the \alf frequency as $k\,L_D>1$. Given that Hall diffusion operates over a subparsec scale in the presence of not very small grains [Eq.~\ref{eq:dhs2}], we see from Eq.~(\ref{eq:wlr}) that the long--wavelength right circularly polarized whistler will be the dominant mode in clouds bereft of small grains. In the presence of very small grains, however, when the Hall scale is much smaller than an astronomical unit, both the whistler and dust cyclotron have similar frequencies. 

In Fig.~\ref{fig:Fx1}(a) we sketch the dispersion curves for the ideal and Hall MHD when $|Z|=1$. The thick solid line in the figure corresponds to the \alf normal mode $\omega=i\,\omega_A$. In the presence of Hall, however, the \alf line splits into whistler (dotted) and dust cyclotron (dashed) branches. At $k\,L_D=1$ \alf degeneracy is lifted. Unlike the Hall MHD of fully or partially ionized dust--free plasmas, where such a splitting of the \alf curve is permanent, in a partially ionized dusty fluid, owing to its dependence on the charge number $|Z|$ on the dust, the branches can move with $|Z|$. For example, these branches can come closer with increasing $|Z|$. This is due to the shrinkage of the Hall scale $L_D$ with $|Z|$.  As a result of the lifting of degeneracy is delayed with increasing $|Z|$ [Fig.~\ref{fig:Fx1}(b)]. 

\begin{figure}
\includegraphics[scale=0.45]{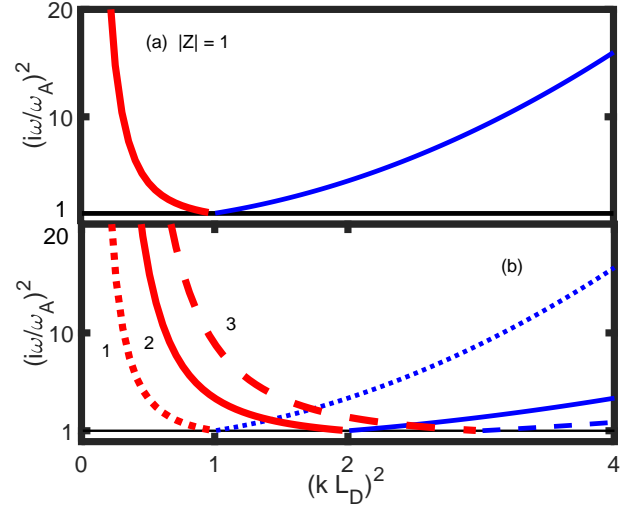}
     \caption{Same as in Fig.~(\ref{fig:Fx1}) but for $(i\,\omega/\omega_A)^2$ against $k\,L_D$ for $|Z|=1$ (a) and $|Z|=1\,,2\,,3$ (b).}
 \label{fig:Fx2}  
\end{figure}
We see from Fig.~\ref{fig:Fx2}(a) that for $|Z|=1$, when $k\,L_D<1$, the low--frequency dust cyclotron is the dominant mode in the system, while the high--frequency whistler is dominant when $k\,L_D>1$. The \alf mode corresponds to $k\,L_D=1$ in the figure. With increasing $|Z|$ [Fig.~\ref{fig:Fx2})(b)], the difference between the low-- and high--frequency branches shrinks. This is also seen in Fig.~\ref{fig:Fx1}(b). 

\section{KH Instability}
The linear analysis in the previous section suggests that the compressibility of the fluid has no bearing on the transverse Hall modes. Thus we shall assume that the fluid is incompressible and the surface of discontinuity in an incompressible, magnetized planar flow exists across the interface $z = 0$.  The flow velocity $v_x (z)$ is assumed to have the following profile:  
\bq
v_x(z) = \left\{
\begin{array}{ll}
v & \mbox{if $z  > 0$}\,,\\
-v & \mbox{if $z < 0$}\,. 
\end{array}
\right.
\label{eq:fpl}
\eq 
The mass density $\rho$ has the same value across the interface. A uniform magnetic field $B$ parallel to the $x$ axis is assumed. 

Given that
\bq
\left(\partial_t  + v\,\partial_x\right) \delta z = \delta v_z\,,
\eq
the continuity of $\delta z$ across the layer gives  
\bq
\Bigg[\frac{\delta v_z}{\left(\omega+i\,k\,v\right)}
\Bigg]=0\,
\label{eq:jcc}
\eq
where [f] denotes the jump in $f$ across the surface of discontinuity.  Defining the Doppler--shifted frequency 
\bq
\sg = \omega + i\,k\,v\,,
\eq
Eq.~(\ref{eq:jcc}) can be written as
\bq
\sg_2\,\delta v_{z1} = \sg_1\,\delta v_{z2}\,.
\label{eq:jc1}
\eq 
 From the momentum Eq.~(\ref{eq:Lm}) we have 
\bq
\left(D^2-k^2\right)\Bigg[\delta v_{z}-\frac{i\,\va^2}{\sg\,B^2}\left(\k\cdot\B\right)\, \delta B_{z}
\Bigg]=0\,,
\label{eq:jm1}
\eq
where $D=d/dz$. Similarly, the induction Eq.~(\ref{eq:Lnd}) gives
\begin{eqnarray}
\left(D^2-k^2\right) \Bigg[\left\{
\left(1+ \frac{\left(\k\cdot\va\right)^2}{\sigma^2}\right)^2
\right. \Bigg.
\nonumber\\
\Bigg.
\left.
-\frac{\left(\k\cdot\B\right)^2}{\sigma^2\,B^2}\,\eta_H^2\,\left(D^2-k^2\right)
\right\}\delta B_z\Bigg]=0\,.
\label{eq:jnd}
\end{eqnarray}
Assuming the solutions of the form
\bq
\delta B_{zj}\,,\delta v_{zj}=\left(A_j\,,C_j\right)\,e^{\pm k\,z} + \left(B_j\,,D_j\right)\,e^{\pm q_j\,z}\,,
\eq
with positive and negative signs for $z>0\,(j=1)$ and $z<0\,(j=2)$, respectively, Eq.~(\ref{eq:jnd}) becomes
\bq
\left(\frac{q}{k}\right)^2 = 1 + \Big[\frac{\left({\sg}^2+ {\omega_A}^2\right)}{\sigma\,k^2\,\eta_H}\Big]^2\,, 
\label{eq:qbk}
\eq
where we have dropped the subscripts $1$ and $2$. From Eqs.~(\ref{eq:jm1}) and (\ref{eq:jnd}) we have
\bq
C_j=\left(\frac{\sg_j}{i\,\left(\k\cdot\B\right)}\right)\,A_j\,,
\nonumber\\
D_j= \left(\frac{i\,\left(\k\cdot\B\right)\,\va^2}{B^2\,\sg_j}\right)B_j\,.
\eq
Using the above equation in Eq.~(\ref{eq:jc1}), which is $\sg_2\left(C_1+D_1\right)= \sg_1\left(C_2+D_2\right)$, yields
\bq
A_1-\frac{\left(\k\cdot\B\right)^2\,v_{A1}^2}{B^2\,\sg_1^2}\,B_1= A_2-\frac{\left(\k\cdot\B\right)^2\,v_{A2}^2}{B^2\,\sg_2^2}\,B_2\,.
\label{eq:BC1}
\eq
The continuity of $\delta B_z$ i.e., $[\delta B_z]=0$, yields
\bq
A_1+B_1=A_2+B_2\,.
\label{eq:BC2}
\eq
The integration of Eq.~(\ref{eq:Lnd}) gives $[\delta J_z]=0$, which yields
\bq
k\,\left(A_1+A_2\right)+ q_1\,B_1 + q_2\,B_2 = 0\,.
\label{eq:BC3}
\eq
The continuity of the pressure across the layer, $[\delta p]=0$ yeilds
\bq
\sg_1\left(C_1+
\frac{q_1}{k}\,D_1\right) + \sg_2\left(C_2+
\frac{q_2}{k}\,D_2\right)=0\,.
\label{eq:BC4}
\eq
From Eqs.~(\ref{eq:BC1})—-(\ref{eq:BC4}) one arrives at the following dispersion relation: 
\begin{eqnarray}
\left({\sg_2}^2-{\sg_1}^2\right)^2\omega_A^2 + \left(2{\omega_A}^2 + {\sg_2}^2 + {\sg_1}^2\right)\left(\frac{q_1}{k}\right)\times 
\nonumber\\
{\sg_1}^2\Big\{
\left({\sg_2}^2+{\omega_A}^2\right)
+ \left(\frac{q_2{\sg_2}^2}{q_1{\sg_1}^2}\right)\left({\sg_1}^2+ {\omega_A}^2\right)
\Big\} = 0\,.
\label{eq:DR}
\end{eqnarray}
Note that the $q \rightarrow \infty$ limit corresponds to the absence of Hall term in the induction equation. The above expression can be simplified by noting that the magnetic field evolves under the combined influence of the fluid advection and Hall diffusion ($\v+\vB$). Thus $q/k$ can be approximated in various limits. For example in the weak diffusion limit, when the fluid advection dominates the Hall diffusion, i.e, when the Hall diffusion time is long (compared to the time over which wave is sheared by advection), 
\bq
\left({\sg}^2+ {\omega_A}^2\right) \gtrsim \sigma\,k^2\,\eta_H\,,
\label{eq:WDL}
\eq
which can also be written as 
\bq
\left(\frac{\sigma}{\omega_A}\right) + \left(\frac{\omega_A}{\sigma}\right) \gtrsim k\,L_D\,.
\eq
This provides the lower bound on the whistler ($\omega_A \ll \sigma$) and dust cyclotron ($\sigma \ll \omega_A$) frequencies. In the weak diffusion limit we may approximate $q/k$ as
 \bq
\left(\frac{q}{k}\right) \simeq \frac{\left({\sg}^2+ {\omega_A}^2\right)}{\sigma\,k^2\,\eta_H}\,. 
\eq
We shall assume that the shear flow profile is given by Eq.~(\ref{eq:fpl}), and analyze the dispersion relation, Eq.~(\ref{eq:DR}) first in the weak diffusion limit. In this limit the dispersion relation reduces to the following simple form: 
\begin{eqnarray} \left(\frac{\omega}{\omega_A}\right)^6+\left(M_A^2+3\right) \left(\frac{\omega}{\omega_A}\right)^4-\big[\left(M_A^2+1\right)^2-4\big]  
\nonumber\\
\times\left(\frac{\omega}{\omega_A}\right)^2-4\, \left(k\,L_D\right)M_A^2\left(\frac{\omega}{\omega_A}\right)-–\left(M_A^2-1\right)^3=0\,, 
\nonumber\\
\label{eq:DRx}
\end{eqnarray}
where the \alf—-Mach number $M_A$ is defined as
\bq
M_A=\left(\frac{v}{\va}\right)\,.
\eq  
In the long--wavelength, low--frequency limit, balancing the dominant first and last terms gives the usual KH mode
\bq
\left(\frac{\omega}{\omega_A}\right)^2 \approx \left(M_A^2-1\right)\,.
\label{eq:kgr}
\eq
Thus in the absence of Hall only super--\alfc ($M_A^2 > 1$) flows cause instability. Note that the low--frequency, long--wavelength limit is the ideal MHD limit when the magnetic field is frozen--in in the partially ionized ideal dusty fluid. However, if the Hall term is retained in the dispersion relation, Eq.~(\ref{eq:DRx}), as a small perturbative correction, even for $M_A = 1$, the wave becomes KH--unstable with the growth rate 
\bq
\left(\frac{\omega}{\omega_A}\right) \approx \left(k\,L_D\right)^{1/3}\propto \left(\frac{|Z|\,n_d}{n_n}\right)^{-1/3}\,
\eq 
suggesting that the instability will be quenched with increasing dust charge density $|Z|\,n_d$.  The inverse dependence on the dust charge density is a typical feature of the dust whistler frequency (Eq.~\ref{eq:wlr}). Therefore, in the presence of \alfc ($M_A = 1$) or super--\alfc ($M_A>1$) flows, in a weakly ionized dusty medium which is dominated by submicron--sized grains, the dust whistler mode becomes unstable.      

\begin{figure}
\includegraphics[scale=0.45]{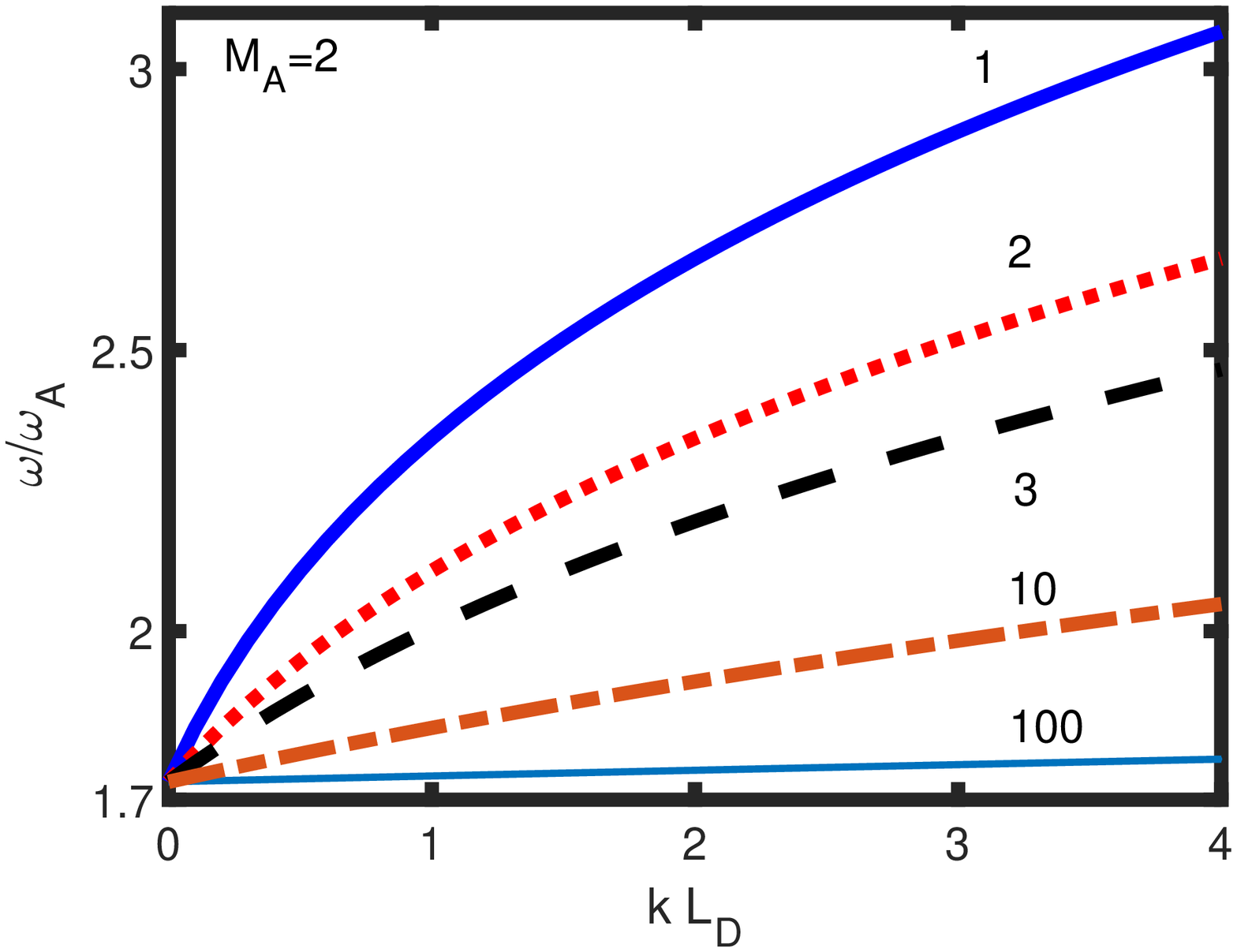}
\includegraphics[scale=0.45]{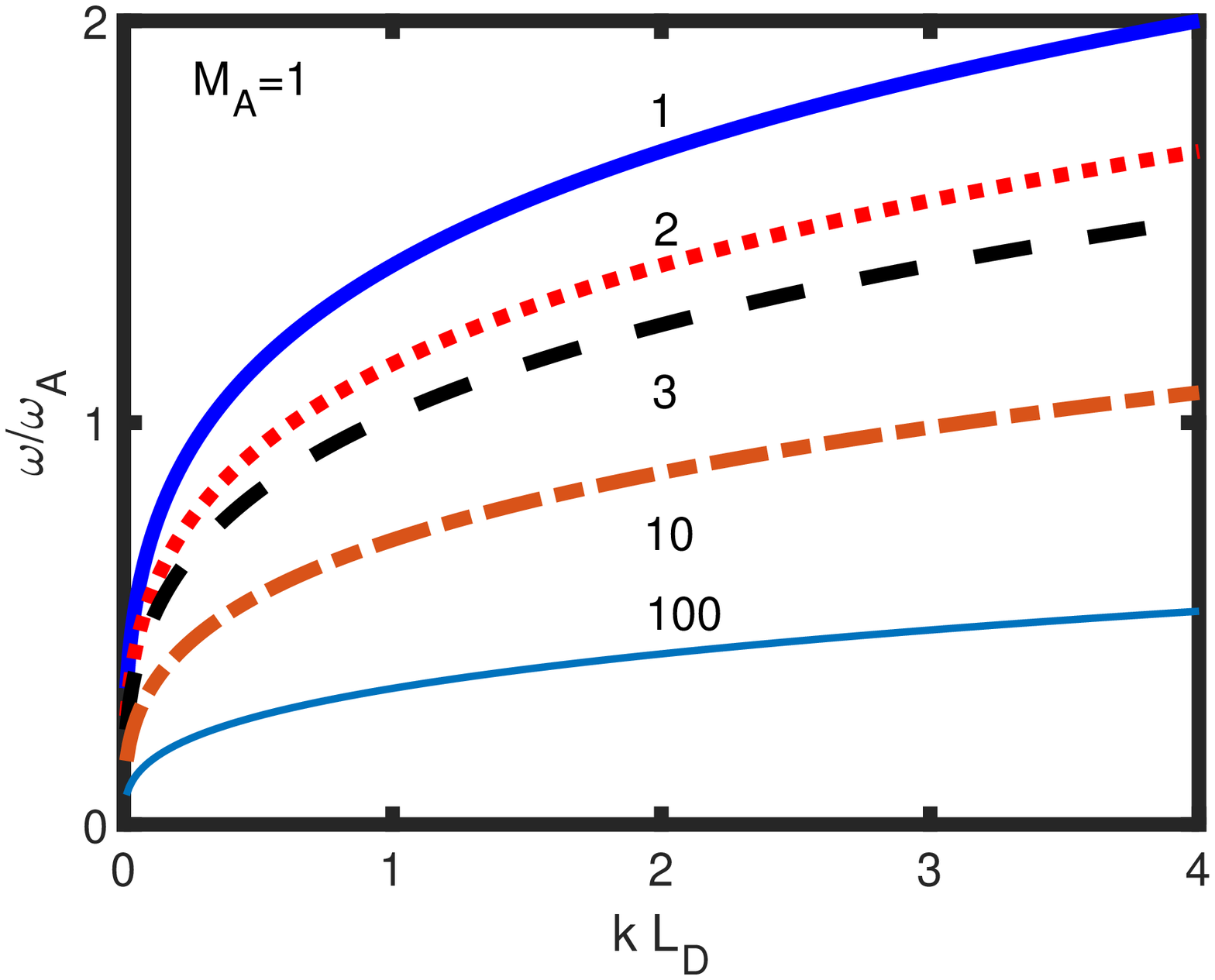}
     \caption{The growth rate (in the units of \alf frequency) against $k\,L_D$ for \alf Mach numbers $M_A = 2$ and $1$ is shown in the above figure for varying $|Z|$, whose values are indicated near the curves.}
\label{fig:fg1}  
\end{figure}

We solve the dispersion relation, Eq.~(\ref{eq:DRx}) and plot the result in Fig.~(\ref{fig:fg1}) for the \alfc and super--\alfc flows. The growth rate of purely growing KH mode further (with respect to the ideal MHD) increases in the presence of charged grains. However, the growth rate falls back to the ideal MHD level with increasing $|Z|$ because Hall diffusion is mitigated as $\sim 1/|Z|$. For very large $|Z|$ Hall is completely quenched, and we are in the ideal MHD regime (corresponding to $|Z|=100$ in the figure). Further increase in $|Z|$ has no bearing on the growth rate.     

The nature of dust whistler and that of dust cyclotron waves are quite different. While whistler waves, like \alf waves, are caused by a balance between the fluid inertia and magnetic tension force, the dust cyclotron waves, are of electrostatic origin and nature.  As can be seen from Eqs.~(\ref{eq:wlr}) and (\ref{eq:icw}), with the increase of charge number density $|Z|\,n_d/n_n$ on the dust, i.e., with a decreasing Hall scale $L_D$, the low--frequency left circularly polarized electrostatic wave may become the dominant normal mode in the fluid.  Given that for the KH mode $k\,v \sim \omega$, the whistler and dust cyclotron exist in two distinct parameter windows corresponding to $M_A \gg 1$ and $M_A \ll 1$, respectively. Therefore, the value of the charge density $|Z|\,n_d/n_n$ or the size distribution of grains determines whether sub-- or super—-\alfc flow will cause the instability in a weakly ionized dusty fluid. 

As the low--frequency ($\omega \ll \omega_A$) limit implies $M_A \ll 1$, balancing the dominant Hall term with the last term in Eq.~(\ref{eq:DRx}) yields 
\bq
\omega \approx \left(\frac{\rho_d}{\rho}\right)\frac{\omega_{cd}}{4\,M_A^2} \propto \frac{|Z|\,n_d}{n_n}\,. 
\label{eq:grl}
\eq
The linear dependence on $|Z|\,n_d/n_n$ is a typical feature of dust cyclotron waves (Eq.~\ref{eq:icw}). Clearly, the presence of charged dust opens up a new channel through which the sub--\alfc flow energy is fed to the waves in a partially ionized dusty fluid.

\begin{figure}
\includegraphics[scale=0.45]{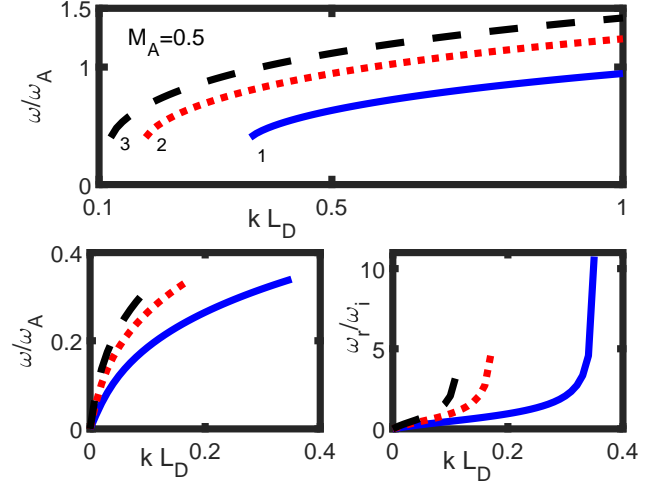}
     \caption{The ratio of the real and imaginary parts of the frequencies, $\omega_r/\omega_i$ against $k\,L_D$ is shown for the overstable modes of Fig.~(\ref{fig:fg1}). The labels by the curves are for the various $|Z|$ values.}
\label{fig:fgx2}  
\end{figure} 
In the sub--\alfc case [Fig.~(\ref{fig:fgx2})], the instability, much like in the dustless case \citep{P18} is entirely due to the Hall effect, although a purely growing mode (top panel) appears beyond a certain $k\,L_D$ cutoff. Note that with decreasing $|Z|$, increasingly shorter wavelength fluctuations are KH--unstable. Further the growth rate of the KH waves also decreases with decreasing $|Z|$. The linear dependence of fluctuation wavelength on $|Z|$ is a typical feature of low--frequency KH instability (Eq.~\ref{eq:grl}). As expected, we see in the top panel that the growth rate increases with increasing $|Z|$. 

We also note the presence of a less rapidly growing overstable mode (lower left panel) in Fig.~(\ref{fig:fgx2}) with the growth rate about one--third of the purely growing (top panel) KH mode. The presence of this mode is generic to the Hall MHD flows \citep{P18}. The overstable mode manifests a dependence on $|Z|$ similar to that of its purely growing counterpart. The ratio of real and imaginary frequencies is shown in the right lower panel of Fig.~(\ref{fig:fgx2}). Notice that with  increasing $k\,L_D$ (i.e. with  decreasing wavelength) the real part of the frequency increases. This implies that the Hall diffusion channels the shear flow energy more efficiently at shorter (with respect to the Hall scale) wavelengths than at longer (ideal MHD) wavelengths.
 
In the highly diffusive limit, long--wavelength $k\,L_D \ll 1$  waves are excluded \citep{P18}. The growth rate of the KH instability for various \alf Mach numbers in the highly diffusive limit is
\bq
\left(\frac{\omega}{\omega_A}\right)
\approx
\left\{
\begin{array}{ll}
\sqrt{M_A} & \mbox{if}\,M_A\ll 1\,,\\
1        &  \mbox{if}\,M_A=1\,,\\
M_A^{3/2}  &  \mbox{if}\,M_A\gg1\,. 
\end{array}
\right.
\eq   
Thus the growth rate of the KH instability is similar to that in the dustless case \citep{P18} except now, the limit $k\,L_D\gg1$ depends on the charge number $|Z|$ on the dust. We may also rewrite the above expressions as
\bq
\left(\frac{\omega}{\omega_{cd}}\right)
\approx
\left\{
\begin{array}{ll}
\sqrt{M_A}\,k\,L_D & \mbox{if}\,M_A\ll 1\,,\\
\,k\,L_D        &  \mbox{if}\,M_A=1\,,\\
M_A^{3/2}\,k\,L_D  &  \mbox{if}\,M_A\gg1\,, 
\end{array}
\right.
\eq   
from where we see that the growth rate is much higher than the dust cyclotron frequency, $\omega_{cd}$. Recall that in the $k\,\L_D\gg 1$ limit, the dust whistler, Eq.~(\ref{eq:wlr}), is the dominant normal mode in the fluid. Therefore, in the highly diffusive limit, the shear flow destabilizes right circularly polarized whistler waves.

\section{Discussion}
Although only $1\,\%$ of mass in molecular clouds is in dust, its presence has a profound effect on magnetic diffusion. The Hall effect may appear due to the relative drift between the magnetized plasma particles and unmagnetized charged dust. In diffuse clouds the plasma number density dominates the negative--charge number density, whereas in dense clouds ($\gtrsim 10^{10}\,\mbox{cm}^{-3}$) most of the negative charge can be soaked by the dust \citep{WN99}. Thus, the presence of grains and the ensuing Hall diffusion will have profound implications in the 
Complex--structure formation in both diffuse and dense cloud cores. This may happen via an interplay between the fluid flow and Hall diffusion of the magnetic field. It appears that even when Hall is the dominant (compared to ambipolar) diffusion mechanism, it may not be dynamically important to regulate the angular momentum transport and ensuing disk formation in collapsing clouds \citep{Z18}. However, explosive growth of the KH instability in the presence of Hall diffusion may cause the formation of filamentary structures in such clouds.

The width of observed molecular spectral lines is attributed to the presence of large-scale supersonic motions in molecular clouds.
These supersonic flows have been attributed to the MHD flows \citep{AM75, MP95}. As circularly polarized waves are the most slowly decaying waves \citep{ZJ83}, it is plausible that these waves are responsible for the observed line broadening. By equating the velocity dispersion measured for molecular lines $(0.3–-5.0)\,\mbox{km}\,\mbox{s}^{-1}$ to the group velocity, one can calculate the wavelength of waves for various regions of the cloud. 

The group velocity of the wave, Eq.~(\ref{eq:whist}) is
\begin{eqnarray}
\left(\begin{array}{c}v_{R}\\v_{L}\end{array}\right) 
= k\,\eta_H
%\times
%\nonumber\\
\Bigg[\frac{1 + \left[\begin{array}{c}2\\6\end{array}\right] 
\frac{1}{k^2\,L_D^2} 
+ \sqrt{1+\frac{4}{k^2\,L_D^2}}
}
{\sqrt{1+\frac{4}{ k^2\,L_D^2}}}
\Bigg]\,.
\label{eq:gv}
\end{eqnarray}
where the subscript $L$ and $R$ denote the left and right polarization. The group velocity of right and left circularly polarized waves differ, albeit by a small number. For definiteness, we choose the group velocity of the right polarized dust whistler waves in the Eq.~(\ref{eq:gv}). Solving for the wavelength $\lambda = 2\,\pi/k$, we get the following cubic equation:
\begin{eqnarray}
\left(\frac{{v_{R}}^2}{{\va}^2} - 1\right)\,\left(\frac{\lambda}{L_D}\right)^3 -– 4\,\pi\,\left(\frac{v_{R}}{\va}\right)\left(\frac{\lambda}{L_D}\right)^2 
\nonumber \\
+\pi^2\,\left(\frac{{v_{R}}^2}{{\va}^2}\right) \left(\frac{\lambda}{L_D}\right)
–- 4\,\pi^3\,\left(\frac{v_{R}}{\va}\right)
= 0\,.
\label{eq:wav}
\end{eqnarray}
Making use of the scaling relation, Eq.~(\ref{eq:fscl}), one gets $0.01\,mG$ for $n_2=1$, which is similar to the one inferred from the starlight polarization data \citep{C11}. For such a field, assuming $|Z|=0.1\,$ and $|Z|=0.001$ for $0.1\,\mu m$ and $.0001\,\mu m$, respectively, the dust cyclotron frequency becomes
\bq
\omega_{cd}=
\left\{
\begin{array}{ll}
 1.3\times 10^{-12}\,B_{-5}\,T_{30}\,a_{-1}^{-3}\,\mbox{s}^{-1}\,,
\nonumber\\
4.72\times 10^{-6}\,B_{-5}\,T_{30}\,a_{.0003}^{-3}\,\mbox{s}^{-1}\,,
\end{array}
\right.
\label{eq:omc}
\eq
where we use $m_d=4\,\pi\,a^3$ for a $3\,\mbox{g}\,\cc$ mean dust material density. For the number densities of diffuse regions and filaments \citep{C11} we get $L_D \lesssim 0.01-1\,\mbox{pc}$ from Eq.~(\ref{eq:dhs2}). Given that $\lambda/L_D\lesssim 1$, according to the leading order from Eq.~(\ref{eq:wav}) we get $\lambda \lesssim 0.01-1\,\mbox{pc}$. In the presence of very small grains, however, the Hall scale is $L_D \sim .001 \,\mbox{au}$ and $\lambda \lesssim .001\,\mbox{au}$. The observations of dense molecular cloud cores suggest the presence of a larger grain size distribution \citep{C89, V93}. Therefore, it is quite likely that subparsec and parsec--scale polarized whistler waves propagate in dense cloud cores and clumps.

As clouds are weakly ionized, ambipolar diffusion may affect wave propagation in the medium. The very survival of the parsec--scale waves requires that the frequency of these waves exceed the neutral--ion collision frequency \citep{KP69, PW08, M11}, or the fluctuation wavelength 
\bq
\lambda \lesssim 2\,\pi\,\va/\nu_{ni} \sim .01\,X_{e-4}^{-1}\,n_{4}^{-1}\,\mbox{pc}\,.
\label{eq:ambd}
\eq  
Here we use the values of $\nu_{in}$ from Eq.~(\ref{eq:cf2}) and fractional ionization from Eq.~(\ref{eq:fion}) after noting that $\nu_{ni}\approx X_e\,\nu_{in}$. Therefore, polarized waves of subparsec  or shorter wavelength may propagate in  molecular clouds ($n_n\sim 10^{2}-—10^{3}\,\mbox{cm}^{-3}$), clumps ($n_n\sim 10^{3}-—10^{4}\,\mbox{cm}^{-3}$) or cores ($n_n\sim 10^{4}-—10^{5}\,\mbox{cm}^{-3}$; \cite{BT07}) without significant damping. In denser cores ($n_n> 10^{6}\,\mbox{cm}^{-3}$) however, ambipolar diffusion will cause significant damping of large wavelength fluctuations.       

The presence of such large--scale fluctuations may help the clumping of the medium and assist in redistributing the magnetic flux. From the induction Eq.~(\ref{eq:ind}), we see that the magneto—-vorticity flux,  
\bq
\frac{d}{d\,t}\int{ \Big[\left(\frac{B}{B_0}\right) + \left(\frac{\rho}{\rho_d}\right)\, \frac{\curl\v}{\omega_{cd}}\Big]\cdot d{\bf{S}}} = 0\,,
\label{eq:flx1}
\eq
is frozen inside a closed surface $S$. Here $B_0$ is some fixed reference value of the magnetic field. Note that depending on the ambient density, the magnetic field and vorticity can be correlated or anticorrelated. From Eq.~(\ref{eq:flx1}) $\curl v \lesssim v / L_D$ and thus for $B/B_0 \sim 1$ we get $v \lesssim \va$. Clearly, the ambient magnetic field in the cloud may induce large--scale radial flow $\lesssim \mbox{km}\, \mbox{s}^{-1}$ which may be responsible for the KH instability. 

The magnetically subcritical elongated structures called striations are an ideal probe with which to investigate the early stages of star formation.  The presence of a magnetic field may not only facilitate the formation of dense filamentary structures along the field lines but may also act as a guiding channel for sub-\alfc flows \citep{L13}.  Magnetically aligned velocity anisotropy appears in sub-\alfc flows in MHD simulations when local thermal pressure is smaller than the magnetic pressure \citep{H08}. Therefore, it has been proposed that the striations are a result of either the KH instability or the magnetosonic waves. If the KH instability is behind the observed diffuse structures, then the maximum growth rate of the KH instability for sub-\alfc flow can be estimated from 
\bq
\left(\frac{\omega_m}{\omega_{cd}}\right) = -\frac{\left(M_A^2-1\right)^2\big[-\left(M_A^2+3\right)+4\left(M_A^2-1\right)^2\big]}{16\,M_A^2}\,,
\label{eq:mrt}
\eq
which except for the corrected sign before $\left(M_A^2+3\right)$, is the same as Eq.~(40) of \cite{P18} if we replace their $\omega_H$ with $\omega_{cd}$. For $M_A=0.5$ \citep{HB12}, the maximum growth rate of the instability becomes $0.14\,\omega_{cd}$ which gives $t_{\mbox{KHI}}= 1\,\mbox{Myr}$ when the dust size distribution is dominated by $0.1\,\mu m$ grains. This is much shorter than the cloud lifetime, which is several megayears \citep{K96}. For very small, PAH grains $t_{\mbox{KHI}}=3 \mbox{yr}$. Clearly, in the presence of very small grains the KH instability grows almost instantaneously. As the wavelength of this very fast growing KH mode is $10^{-3}\,AU$ it would appear that the small grains are dynamically unimportant to the large--scale structure formation. However, dust in molecular clouds interacts with gas, metals, and dust particles. In dense clouds, dust grains grow their size by accretion onto grain mantles and coagulation. Owing to the high 
surface—to--volume ratios, accretion has a predominant influence on the small ($a<0.03\,\mu m$) grains \citep{H15}. The accretion hardly affects the grain size distribution when $a>0.03\,\mu m$. Coagulation, i.e. grain-—grain sticking (due to collision), on the other hand, converts small grains into large grains \citep{C93, DT97}. 

The typical accretion time is \citep{HV14}
\bq
\tau_{\mbox{acc}}\simeq 10^8\,a_{0.1}\,n_{3}^{-1}\,T_{30}^{-1/2}\,\left(\frac{Z}{Z_{\sun}}\right)^{-1}\left(\frac{S_{\mbox{acc}}}{0.3}\right)^{-1}\,\mbox{yr}\,,
\eq
where $Z$ is the metallicity and $S_{\mbox{acc}}$ is the sticking probability for accretion. The coagulation time depends on the dust velocity dispersion $v$ and is \citep{A17}
\bq
\tau_{\mbox{coag}}\simeq 10^8\,a_{0.1}\,n_{3}^{-1}\,v_{.01}^{-1}\left(\frac{\rho_d/\rho_n}{.01}\right)^{-1}\,\mbox{yr}\,.
\eq
Here we assume that the dust material density is $3\,\mbox{g}\,\mbox{cm}^{-3}$ and $v=0.01\,\mbox{km}\,\mbox{s}^{-1}$. 
Therefore, both grain accretion and grain coagulation operates over similar timescales. Thus in the presence of $0.1 \mu \mbox{m}$ grains the $t_{\mbox{KHI}}= 1\,\mbox{Myr}$ is much shorter than both the accretion and coagulation time scales. Note that large grains are destroyed due to grain-—grain collision \citep{Y04} and shock waves \citep{M89a}. However, these processes are slow: the collisional destruction time is $54\,\mbox{Myr}$ \citep{A17}, whereas the destruction time with shock waves is  $4\times 10^2\,\mbox{Myr}$ \citep{M89a}. Therefore, the destruction of grains will have no bearing on the KH instability in cloud cores dominated by large grains.

In dense clouds ($n_6\gtrsim 1$) as 
$\tau_{\mbox{acc}} \sim \tau_{\mbox{coag}} \ll t_{\mbox{KHI}}$, small grains are dynamically processed into large grains long before the onset of the KH instability. Therefore, accretion and coagulation are important in dense clouds, whereas  these processes operate over a much longer (than the KH) timescale in diffuse clouds and filaments. The wavenumber corresponding to the maximum growth rate is 
\bq
k_m\,L_D = \frac{\left(M_A^2-1\right)^3\big[-\left(M_A^2+3\right)+4\left(M_A^2-1\right)^2\big]}{128\,M_A^4}\,.
\label{eq:wln}
\eq
Thus the wavelength corresponding to the maximum growth rate is $\lambda_m \sim 7.5\,L_D$ which for submicron--sized grains [Eq.~\ref{eq:dhs2}], is $\gtrsim 0.1 \mbox{pc}$ in dense cloud cores. The projected wavelength of the most common striations in the envelop of the Taurus molecular cloud is $0.23\,\mbox{pc}$ \citep{H16}. Therefore, it is plausible that the KH instability has some role in forming striations. However, given the uncertainty of the physical parameters and the limitations of the linear model, the role of the KH instability in structure formation is only suggestive and needs further investigation.  

\section{summary} 
The KH instability in a partially ionized and magnetized dusty fluid, depending on the dust charge density or the dust size distribution, can be analytically studied in the weakly and highly diffusive limits. In the weak diffusion limit, i.e. when the Hall diffusion time is comparable to or longer than the time over which the wave is sheared, the growth rate of the instability in the presence of sub-\alfc flow increases with $|Z|$ on the grain, while it is quenched in the presence of \alfc or super-\alfc flow. On the other hand, in the highly diffusive limit, the growth rate of the instability only indirectly depends on the dust charge. The instability in this case grows linearly with the \alf frequency; the slope of the curve is determined by $M_A^{\alpha}$ where $\alpha=0.5\,,0\,,1.5$ for the sub--\alfc, \alfc and super--\alfc flows, respectively.  

To summarize, (1) In the absence of small grains, Hall diffusion operates over subparsec and parsec scales in dense and diffuse clouds, respectively.\\    
(2) The polarization of waves in a partially ionized dusty medium depends on the dust charge density or the grain size distribution. In clouds bereft of small grains, right circularly polarized whistler is the dominant normal mode, whereas in the presence of very small grains both whistler and dust cyclotron have similar frequencies. 
\\ 
(3) In the presence of shear flows, these waves may become KH--unstable with the charge number on the grain or the grain size distribution operating as a switch to the instability.\\
(4) When Hall diffusion is weak, i.e. diffusion time is comparable to or longer than the advection time, the growth rate of the KH instability may increase or decrease depending on the charge number on the grain. In super--\alfc flows the instability grows at a faster rate than without Hall.\\
(5) In this weak diffusion limit, a new overstable mode whose growth rate is lower than that of the purely growing KH mode appears in the dusty fluid.\\ 
(6) If the Hall diffusion time is shorter than the advection time, the 
shear flow destabilizes the right circularly polarized whistler waves with the growth rate much higher than the dust cyclotron frequency.
\vspace{4mm}
{\begin{center}{\bf Acknowledgements}\end{center}}
BP wishes to thank Prof. Mark Wardle for his many insightful comments on this subject. 

\end{document}